\documentclass{article}

\newcommand{\DOWNUP}{\mbox{\boldmath
     $\mbox{}_{\displaystyle \sqcup}\hspace{-1.0 mm}\mbox{}^{\displaystyle
     \sqcap}$}}
\newcommand{\DOWNUPDOWN}{\mbox{\boldmath
     $\mbox{}_{\displaystyle \sqcup}\hspace{-1.0 mm}
     \mbox{}^{\displaystyle \sqcap}\hspace{-1.0 mm}
     \mbox{}_{\displaystyle \sqcup}$}}

\def\Reell{{\rm I\!R}}

\begin{document}

\title{The relation of the Allan- and $\Delta$-variance to the continuous
  wavelet transform} 

\author{M. Zielinsky\thanks{I. Physikalisches Institut, Universit\"at
    zu  K\"oln,   Z\"ulpicherstr. 77, D-50937 K\"oln, Germany, email: {\em lastname}@ph1.uni-koeln.de } \and J. Stutzki$^{*}$}

\maketitle

\begin{abstract}
This paper is understood as a supplement to
the paper by [Stutzki et al, 1998], where we have shown the
usefulness of the Allan-variance and its higher dimensional
generalization, the $\Delta$-variance, for the characterization of
molecular  cloud structures.  In this study we
present the connection 
between the Allan- and $\Delta$-variance and a more popular
structure analysis tool: the wavelet transform. We show that the  Allan- and
$\Delta$-variances are the variances of wavelet transform
coefficients. 

\end{abstract}

\section{Introduction}
In \cite{stutzki} (hereafter paper I) we defined the $\Delta$-variance as a generalization of
the Allan-variance, a method traditionally used in the stability
analysis of electronic devices. The Allan- or $\Delta$-variance
method allows a clear separation between the global drift characteristics 
of the signal and other  contributions, like white noise. It is specially 
suited to estimate the spectral index of power law spectral distributions. 
We demonstrated the use of the two-dimensional $\Delta$-variance
 for the analysis of molecular cloud
images, e.g. maps of CO line emission. Application of the
$\Delta$-variance analysis to the integrated CO intensity
images of various molecular clouds shows 
a power law behaviour of power spectrum of the intensity distribution
with spectral 
index $\beta$ close to 2.8.  
This intensity distribution  has  a structure 
well described by a {\em fractional Brownian motion} ({\em fBm}) model
and in paper I it was shown that the
power law index $\beta$ derived by the $\Delta$-variance is  related to
the drift exponent $H$ of the corresponding {\em fBm} model by
$\beta=E+2H$ (where $E$ is the Euclidean dimension of the model) and
to other parameters characterizing the fractal, such as the
area-perimeter index.

Another tool well established over the last decade in the areas of
structure analysis and data processing is the wavelet transform. While
the discrete wavelet transform is mainly used in the field of signal
compression, the continuous wavelet transform has applications in
texture analysis and feature extraction.  The connection between the
local regularity of a function and the scaling behaviour of its
wavelet transform coefficients, as the scale considered becomes smaller
and smaller, makes the wavelet technique a natural tool for the
analysis of self-similar distributions with (multi-)fractal properties 
(eg. \cite{holschneider}).

In the context of molecular cloud structure analysis the wave\-let
transform was first used 
by \cite{gill}, who also pointed out the
advantages of this method over traditional structure indicators like
the autocorrelation function. In their study they found a power law
behaviour for the centroid velocities of $^{13}$CO spectral data of
L1551, as well as for the peak intensities.
\cite{langer}, presented a clump decomposition of CO maps of the cloud 
Barnard 5 using the Laplacian pyramid transform, which is also a
wavelet  transform. The number distribution of clump masses derived by 
the Laplacian pyramid transform follows a power law, in agreement with 
the results presented by \cite{kramer}, obtained by using  gaussian clump
decomposition. These results give additional evidence for the 
hierarchical structure of molecular clouds.

The purpose of this paper is to show that both
these concepts are closely related, namely that the Allan- and
$\Delta$-variance can be expressed using the variance of wavelet
transform coefficients for suitably chosen wavelets.  

After a short introduction of the Allan-variance and the wavelet
transform in one dimension we show the relation between
these concepts (section 2). The use of the $\Delta$-variance and the
continuous wavelet transform generalizes our result to two dimensions
(section 3). We summarize our results in section 4.

\section{Allan-variance and wavelets}
\subsection{Allan-variance}
We first repeat some facts about the Allan-variance; for a broader
introduction the reader may consult  appendix A of paper I.

To calculate the Allan-variance of a random signal $s(t)$ we consider
the differences $d(t,T)$ of succeeding averages over a (time) interval $T$
and calculate their variance. 
Defining the {\em down-up rectangle function}  by 
\begin{displaymath}
\DOWNUP_{T}(t) := \left\{ \begin{array}{r@{\quad }l} 
  \frac{-1}{2T} & \mbox{:\quad} -T \le t < 0 \\
   \frac{1}{2T} & \mbox{:\quad} 0 \le t \le T \\
  0 & \mbox{:\quad elsewhere, }  
 \end{array} \right. 
\end{displaymath}
this difference can be written as a convolution  
\begin{eqnarray}
\label{diff}
\frac{1}{T} \left( ~  \int\limits_{t-T}^{t} s(t')dt' - 
\int\limits_{t}^{t+T} s(t')dt' \right) & = & \nonumber\\
-\left\{s * 2 \hspace{.5mm}\DOWNUP_{T}\right\}(t): & = & d(t,T).
\end{eqnarray}
Without loss of generality  we here assume the average $\bar{s}$ to be
zero. Then the Allan-variance $\sigma^{2}_{A}(T)$ is defined as 
\begin{equation}
\label{allandef}
\sigma^{2}_{A}(T):=\frac{1}{2}\langle d^{2}(t,T) \rangle_{t},
\end{equation}
where  $\langle ... \rangle_{t}$ denotes the average
over all times (or positions)  $t$.

\subsection{Wavelets}
The Fourier transform of a function
$\psi$ is defined by $\tilde{\psi}(\omega) =  \int_{\Reell} 
e^{-i 2\pi \omega t} \psi(t) dt$.
A square integrable function  $\psi$, 
which satisfies 
\begin{equation}
\label{wavdef}
0 < c_{\psi}:=  \int_{\Reell}
\frac{| \tilde{\psi}(\omega)|^{2}}{|\omega|} d\omega < \infty \nonumber
\end{equation}
is called a {\em wavelet}. For $a \in \Reell \backslash \{0\}, b \in \Reell$
the wavelet transform of a signal $s$  
with respect to $\psi$ is defined by 
\begin{equation}
\label{wavtrafo}
L_{\psi}s(a,b):= \frac{1}{\sqrt{c_{\psi}}}
|a|^{-1/2} \int_{\Reell} s(t)
\psi\left( \frac{t-b}{a}\right) dt.
\end{equation}
Each wavelet transform
coefficient $L_{\psi}s(a,b)$ is thus the scalar product of the signal with
a dilated and translated version of the  "mother function" $\psi$.

\subsection{The Allan-variance as variance of wavelet transform coefficients} 

As a special case of a mother function  consider a translated and
dilated version of the well known Haar wavelet $\psi_{\tilde{H}}$,  
which we define here  as follows: 
\begin{displaymath}
 \psi_{\tilde{H}}(t) := \left\{ \begin{array}{r@{\quad }l@{\quad }c}
  1 & : & -1 \le t < 0 \\
  -1 &: &  0 \le t \le 1 \\
  0 & :& \mbox{elsewhere}  
 \end{array} \right.. 
\end{displaymath}
For this the constant
defined by  (\ref{wavdef}) is $c_{\psi_{\tilde{H}}}=8 \ln 2$ (see Appendix).
In the case of this modified  Haar wavelet the transform coefficient
$L_{\psi}s(a,b)$  given by 
(\ref{wavtrafo}) corresponds to the  difference of succeeding
integrals over $s(t)$, which is the same as the averaging process leading
to the definition of  the Allan-variance. 
It  follows that 
\begin{displaymath}
\int\limits_{\Reell} \psi_{\tilde{H}}\left( \frac{t-b}{a}\right) s(t) dt=
\int\limits_{b-a}^{b} s(t) dt - 
\int\limits_{b}^{b+a} s(t) dt,
\end{displaymath}
which  equals the term in brackets in eq.
(\ref{diff}) with $b \equiv t$ and $a \equiv T$.  Further we have
\begin{eqnarray}
L_{\psi_{\tilde{H}}}s(a,b) & = &   \frac{1}{\sqrt{c_{\psi_{\tilde{H}}}}}
|a|^{-1/2} \int_{\Reell} s(t)
\psi\left( \frac{t-b}{a}\right) dt \nonumber\\
& = & \frac{1}{\sqrt{c_{\psi_{\tilde{H}}}}}
|a|^{-1/2} \left( ~ \int\limits_{b-a}^{b} s(t) dt - 
\int\limits_{b}^{b+a} s(t) dt \right) \nonumber\\
&\stackrel{\mbox{\tiny(\ref{diff})}}{=}&
\sqrt{\frac{|a|}{c_{\psi_{\tilde{H}}}}} \cdot d(b,a) \nonumber \\
 \mbox{i.e.~~}
d(b,a) & = & \sqrt{\frac{c_{\psi_{\tilde{H}}}}{|a|}} L_{\psi_{\tilde{H}}}s(a,b).
\end{eqnarray}
Using eq. (\ref{allandef}) it follows that the Allan-variance can be
expressed as
\begin{equation}
\label{allanwavelet}
\sigma_{A}^{2}(a)=\frac{1}{2}\frac{c_{\psi_{\tilde{H}}}}{|a|}
\langle L_{\psi_{\tilde{H}}}^{2}s(a,b) \rangle_{b}, 
\end{equation} 
that is the Allan-variance at lag $a$ equals the  average of
the wavelet transform coefficients  over the translation index $b$ for
constant  dilation index $a$.

\section{$\Delta$-Variance and $\Delta$-Wavelet}

The main idea of the proceeding section was, that the difference of
adjacent averages corresponds with
the convolution product of the signal $s$ with an oscillating function
$\psi$ of zero mean, that is a wavelet transform coefficient $L_\psi s$.
 This concept will now be extended  to the
symmetric version  of the Allan-variance, the   
$\Delta$-variance as introduced in paper I. 
We recall from  paper I
(Appendix A, 
p. 713 ff.) the  definition of the one-dimensional {\em
down-up-down rectangle function}
\begin{displaymath}
\DOWNUPDOWN(x) := \left\{ \begin{array}{r@{\quad }l} 
  1 & \mbox{:\quad} |x| \le 1/2 \\
  -1/2 & \mbox{:\quad} 1/2 < |x| \le 3/2 \\
  0 & \mbox{:\quad elsewhere }  
 \end{array} \right. .
\end{displaymath} 
The  {\em double difference} $\Delta(t,T)$, that is the average over
succeeding intervals is defined as follows:
\begin{eqnarray}
\label{doppdiff}
 \Delta(t,T): & = & \left\{s(...)*\frac{1}{T}
\DOWNUPDOWN(\frac{...}{T})\right\}(t) \nonumber\\
 & = & \int\limits_{\Reell} dx
\frac{1}{T} \DOWNUPDOWN\left(\frac{t-x}{T}\right) s(x).
\end{eqnarray}
The $\Delta$-variance is the  variance of the  {\em double
difference} $\Delta(t,T)$:
\begin{equation}
 \label{delta1var}
  \sigma_{\Delta}^{2}(T):=\frac{1}{2}\langle \Delta^{2}(t,T) \rangle_{t}.
\end{equation}
To find the relation to a wavelet transform we choose 
the {\em down-up-down rectangle function}
$\DOWNUPDOWN(x)$ as mother function, which defines the family of 
one-dimensional "$\Delta$-wavelets"\footnote{A scaled version of this
  $\Delta$-wavelet is used as "French Hat" wavelet in the literature,
  cf. \cite{gill}.}: 
\begin{displaymath}
  \psi_{\Delta_{a,b}}(x):= \DOWNUPDOWN\left(\frac{x-b}{a}\right),
\end{displaymath}
where again $a \in \Reell/\{0\}$ and $b \in \Reell$. The value of the
admissibility constant is $c_{\psi_{\Delta}} \approx 3.37$.
The corresponding wavelet transform coefficients are 
\begin{displaymath}
 L_{\psi_{\Delta}}s(a,b)  =  \frac{1}{\sqrt{c_{\psi_{\Delta}}}} |a|^{-1/2}
\int\limits_{\Reell} dx 
\DOWNUPDOWN\left(\frac{x-b}{a}\right) \cdot s(x), 
\end{displaymath}
and with  $a \equiv
T, b \equiv t$ and  $\DOWNUPDOWN(x)=\DOWNUPDOWN(|x|)$ one gets
\begin{eqnarray}
  \Delta(b,a) &\stackrel{\mbox{\tiny(\ref{doppdiff})}}{=}  & \sqrt{\frac{c_{\psi_{\Delta}}}{|a|}}
L_{\psi_{\Delta}}s(a,b). \nonumber 
\end{eqnarray}
With eq. (\ref{delta1var}) this leads to the result analogous to eq.
(\ref{allanwavelet}): 
\begin{equation}
 \label{deltawavelet}
 \sigma_{\Delta}^{2}(a)=\frac{1}{2} \frac{c_{\psi_{\Delta}}}{|a|}\langle
L^{2}_{\psi_{\Delta}}s(a,b)\rangle_{b}.
\end{equation}

\subsection{Two-dimensional Continuous Wavelet Transform}
To generalize our results to higher dimensions we make use of the definition
of two-dimensional (directional) wavelets given by \cite{anto}.  
As in the one-dimensional case a two-dimensional function $\psi$ 
is called a wavelet, if the
following admissibility condition holds:
\begin{displaymath}
0 < c_{\psi}:= \int_{\Reell^{2}}
\frac{| \tilde{\psi}(\vec{f})|^{2}}{\|\vec{f}\|^{2}} d^{2}\vec{f} < \infty.
\end{displaymath}
From  \cite{anto} we adopt the notation for translation, dilation and rotation
operators:\\ 
\vspace*{2mm} \\
$\mbox{translation: \quad}  (T^{\vec{b}}s)(\vec{x})  =
 s(\vec{x}-\vec{b}), ~~b\in \Reell^{2} \\
 \mbox{dilation: \hspace{0.35cm} \quad}  (D^{a}s)(\vec{x})  =
 \frac{1}{a}s(\frac{\vec{x}}{a}), ~~a>0  \\
 \mbox{rotation:\hspace{0.4cm}  \quad} (R^{\theta}s)(\vec{x})  =
s(r_{-\theta} (\vec{x})), ~~\theta\in[0,2\pi),\\$\vspace*{2mm}\\
where $r\in SO(2)$. A combination $\Omega(a,\theta,\vec{b}) =
T^{\vec{b}}D^{a}R^{\theta}$  of these
operators acts on a function $s$ in the following way:
\begin{displaymath}
 (\Omega(a,\theta,\vec{b})s)(\vec{x})=s_{a,\theta,\vec{b}}(\vec{x})
\equiv \frac{1}{a} s\left(\frac{1}{a} r_{-\theta} (\vec{x}-\vec{b})\right).
\end{displaymath}
With this notation the two-dimensional continuous wave\-let transform
(CWT) of a function (or ``image'') $s$  
with respect to  $\psi$ is given by
\begin{equation}
\label{wav2trafo}
L_{\psi}s(a,\theta,\vec{b}):= \frac{|a|^{-1}}{\sqrt{c_{\psi}}}
\int_{\Reell^{2}} d^{2}\vec{x}~ 
\psi\left(r_{-\theta}(\frac{\vec{x}-\vec{b}}{a})\right) s(\vec{x}) 
\end{equation}
with $a \in \Reell \backslash \{0\}, \vec{b} \in \Reell^{2}$.
For a complex valued mother function $\psi$ in eq. (\ref{wav2trafo})  is
replaced by  its  complex 
conjugate  $\psi^{*}$, but we will only make use of real valued
functions here.   
At present only isotropic wavelets $\psi(\vec{r})=\psi(r), r=|\vec{r}|$
will be considered, so that we omit the  index $\theta$: 
\begin{displaymath}
L_{\psi}s(a,\vec{b}):= \frac{|a|^{-1}}{\sqrt{c_{\psi}}}
 \int_{\Reell^{2}} d^{2}\vec{x}~ 
\psi\left( \frac{\vec{x}-\vec{b}}{a}\right) s(\vec{x}). 
\end{displaymath}

\subsection{$\Delta$-Variance in Two Dimensions}
This section is based on paper I, Appendix B (p. 717), where we
now consider the case $E=2$, that is two-dimensional signals or images.

The two-dimensional {\em down-up-down rectangle function} is given by:
\begin{eqnarray}
\label{twodmother} 
 \stackrel{2}{\DOWNUPDOWN}(\vec{r} ) & : = &
 \frac{3^{2}}{(3^{2}-1)}\left[~\sqcap(\vec{r}) -
   \frac{1}{3^{2}}\sqcap(\frac{\vec{r}}{3})~\right] \nonumber\\ & = &  
\left\{ \begin{array}{r@{\quad }l@{\quad }c}
  1    & : &  |\vec{r}| \le 1/2 \nonumber\\
  -1/8 & : & 1/2 < |\vec{r}|  \le 3/2  \\
  0    & : & \mbox{ elsewhere }  
 \end{array} \right.\nonumber. 
\end{eqnarray}
For the two-dimensional unit sphere the "volume" is
${\cal V}_{2}= \pi$ and the "surface" is  ${\cal S}_{2}=2 \pi$. 
Using the nomenclature of paper I we have 
\begin{equation}
\label{twodfilter} 
 \bigodot_{D,2}(\vec{r})=\frac{1}{\pi (\frac{D}{2})^{2}}
\stackrel{2}{\DOWNUPDOWN}(\frac{\vec{r}}{D}). 
\end{equation}
Convolving $s(\vec{r})$ with $\bigodot_{D,2}(\vec{r})$ leads to the
difference of averages at average distance $D$:
\begin{eqnarray}
\label{diff2}
  \Delta_{s}(\vec{r},D,2) & := & \left\{ s(...) *
    \bigodot_{D,2}(...)\right\}(\vec{r})  \nonumber\\
 & = & \int\limits_{\Reell^{2}} d^{2}\vec{x} \bigodot_{D,2}(\vec{r}-\vec{x})
\cdot s(\vec{x}) \nonumber\\
 & =  & \frac{1}{\pi \left(\frac{D}{2}\right)^{2}}
\int\limits_{\Reell^{2}} d^{2}\vec{x} \hspace{.8mm}
\DOWNUPDOWN\left(\frac{\vec{r}-\vec{x}}{D}\right) \cdot s(\vec{x}) 
\end{eqnarray}
As in the one-dimensional case the $\Delta$-variance is given as the
variance, i.e the
autocorrelation function at zero lag of the filtered signal:
\begin{equation}
  \label{sigma2dim}
  \sigma^{2}_{\Delta_{s}}(D):=\frac{1}{2 \pi} A_{\Delta_{s}(\vec{r},D,2)}(0).
\end{equation}

\subsection{$\Delta$-Variance and Two-dimensional Continuous Wave\-let Transform}
To relate the $\Delta$-variance to the two-dimensional CWT one
naturally  considers
the {\em down-up-down-rectangle function} given by eq. (\ref{twodmother})
as mother wavelet and defines   
\begin{displaymath}
  \psi_{\Delta_{a,\vec{b}}}(\vec{r}):=
\stackrel{2}{\DOWNUPDOWN}\left(\frac{\vec{r}-\vec{b}}{a} \right); 
\end{displaymath}
$\psi_{\Delta_{a,\vec{b}}}$ will also be called 
(two-dimensional) {\em $\Delta$-wavelet}; the admissibility
constant for the two-dimensional function is
$c_{{\psi}_{\Delta}} \approx 4.11$.
Like in the one-dimensional case the wavelet transform coefficients 
$L_{\psi_{\Delta}}s(a,\vec{b})$ are directly related to the differences 
of average  values at average distance $D$ $\Delta(\vec{r},D,2)$ (cf. eq.  
(\ref{diff2})):
\begin{eqnarray}
  \label{wavdelta}
  L_{\psi_{\Delta}}s(a,\vec{b}) & = & \frac{1}{\sqrt{c_{\psi_{\Delta}}}}
  \frac{1}{a} \int\limits_{\Reell^{2}} d^{2}\vec{x} \hspace{.8mm}
  \psi_{{\Delta}_{a,\vec{b}}}(\vec{x}) \cdot s(\vec{x}) \nonumber\\
 & = & \frac{1}{\sqrt{c_{\psi_{\Delta}}}}  \frac{1}{a} \int\limits_{\Reell^{2}}
d^{2}\vec{x} \hspace{.8mm} \DOWNUPDOWN
\left(\frac{\vec{x}-\vec{b}}{a}\right) \cdot  s(\vec{x})
\end{eqnarray}
We identify $D \equiv a$ and $\vec{r} \equiv \vec{b}$ and because of
$\DOWNUPDOWN(\vec{r})=\DOWNUPDOWN(|\vec{r}|)$ we find 
using  eq. (\ref{diff2}) and eq. (\ref{wavdelta}):
\begin{eqnarray}
 L_{\psi_{\Delta}}s(a,\vec{b}) & = & \frac{1}{\sqrt{c_{\psi_{\Delta}}}}
  \frac{1}{a} \cdot \frac{\pi a^{2}}{4} \cdot \Delta_{s}(\vec{b},a,2) \nonumber\\
& = & \frac{\pi a}{4 \cdot \sqrt{c_{\psi_{\Delta}}} } \cdot
\Delta_{s}(\vec{b},a,2).  \nonumber
\end{eqnarray}
Looking at the  autocorrelation function as the
two-dimensional  generalization of the variance as we have done in eq. 
(\ref{sigma2dim}), we find as the main result of this section:
\begin{equation}
  \label{mainresult}
     \sigma^{2}_{\Delta_{s}}(a)= \frac{1}{2 \pi} \frac{16 \hspace{.8mm}
       c_{\psi_{\Delta}}}{\pi^{2} a^{2}} \hspace{.8mm}
A_{L_{\psi_{\Delta}}s(a,\vec{b})}(0),
\end{equation}
i.e. the $\Delta$-variance for a given lag $a$ can be expressed in terms
of the autocorrelation function at zero lag of the wavelet transform
coefficients 
with dilation parameter $a$.

\section{Summary and Outlook}
The main results of our study are  equations (\ref{allanwavelet}),
(\ref{deltawavelet}) and 
(\ref{mainresult}) which show, that the Allan- and $\Delta$-variance
can be expressed 
using the coefficients of special wavelet transforms and that both these 
concepts are very closely related.  
 \cite{pando}
have shown that the {\em discrete} wavelet transform is a good power
spectrum estimator especially in the case of finite sized
samples. The same result was found by \cite{bensch} for the $\Delta$-variance
analysis of CO integrated intensity maps as small as $32 \times 32$ pixel.  
Detailed investigations by Bensch et al. and \cite{ossk} show that
the $\Delta$-variance remains a robust measure of astronomical images 
if the effects of noise, smearing due
to the finite size antenna beam pattern and radiative transfer are
taken into account.

Pando et al.  pointed out  that evaluating the power spectrum 
with  an over-complete basis
function system of the CWT can lead to correlations "that are not in 
the sample, but due to correlations among the 
wavelet coefficients". This will also be true for the Allan- and the
$\Delta$-variance, because the definitions of both  employ such an 
over-complete system. 
It has to be checked in a further study, how the results change, 
if only coefficients on a discrete, wavelet adapted grid (a
"wavelet frame") are considered, i.e. if the continuous wavelet
transform is replaced by a discrete one.

While the choice of isotropic wavelets was natural and sufficient in
our work on the 
isotropic scaling behaviour of molecular cloud structure, the directional
sensitivity of non-axial symmetric wavelets will allow the determination of
preferred orientations, eg. of filamentary substructures, in
astronomical images.

\begin{appendix}
\section{Evaluation of the factors $c_{\psi}$}
For all wavelets discussed in this paper the factors 
\begin{displaymath}
c_{\psi}:=  \int_{\Reell}
\frac{| \tilde{\psi}(\omega)|^{2}}{|\omega|} d\omega \mbox{\quad and \quad}
c_{\psi}:=  \int_{\Reell^{2}}
\frac{\| \tilde{\psi}(\vec{f})\|^{2}}{\|\vec{f}\|^{2}} d^{2}\vec{f}
\end{displaymath}
for the one- and two-dimensional case, respectively,
can be calculated analytically. The integrals  were solved using {\em Mathematica 3.0  for students}.

\subsection{Modified Haar-Wavelet}
Looking at the mother function $\psi(x)=-2 \DOWNUP(x)$ of the modified
Haar wavelet system we find (paper I, p.714) 
\begin{displaymath}
| \tilde{\psi}(\omega)|^{2}=4 \frac{\sin^{4}(\pi \omega)}{(\pi
  \omega)^{2}} \mbox{\quad and therefore}
\end{displaymath}
\begin{eqnarray}
c_{\psi_{\tilde{H}}} & = & \int\limits_{- \infty}^{\infty} 4  \frac{\sin^{4}(\pi \omega)}{(\pi
  \omega)^{2} \cdot |\omega|} \nonumber
 =  \int\limits_{- \infty}^{\infty} 4\pi  \frac{\sin^{4}(\pi \omega)}{|\pi
  \omega|^{3}}\nonumber
 =  8\ln 2 \nonumber
\end{eqnarray}

\subsection{$\Delta$-Wavelet}
\subsubsection{$\Delta$-Wavelet in One Dimension}
Looking at the mother function $\psi(x)=\DOWNUPDOWN(x)$ we find (paper
I, p.715) 
\begin{displaymath}
| \tilde{\psi}(\omega)|^{2}=4 \frac{\sin^{6}(\pi \omega)}{(\pi
  \omega)^{2}} \mbox{\quad and therefore}
\end{displaymath}
\begin{eqnarray}
c_{\psi_{\Delta}} & = & \int\limits_{- \infty}^{\infty} 4  \frac{\sin^{6}(\pi \omega)}{(\pi
  \omega)^{2} \cdot |\omega|}\nonumber
 =   \int\limits_{- \infty}^{\infty} 4\pi  \frac{\sin^{6}(\pi \omega)}{|\pi
  \omega|^{3}} \nonumber\\
& = & 12\ln 2 - \frac{9 \ln 3}{2} \approx 3.37 \nonumber
\end{eqnarray}

\subsubsection{$\Delta$-Wavelet in Two Dimensions} 
The power spectrum of the two-dimensional {\em down-up-down rectangle
  function} is (paper I, p. 717)
\begin{displaymath}
  \|\widetilde{\DOWNUPDOWN}(\vec{f})\|^{2}=\left[ \frac{J_{1}(\pi f)}{(\pi f)} -
   \frac{J_{1}(3 \pi f)}{(3 \pi f)}\right]^{2},
\end{displaymath}
where $J_{1}$ denotes the first order {\em Bessel}-function of first
kind. With this we get
\begin{eqnarray}
c_{\psi_{\Delta}} & = & \int\limits_{\Reell^{2}} \frac{\|
    \widetilde{\DOWNUPDOWN}(\vec{f})\|^{2}}{\|\vec{f}\|^{2}} d^{2}\vec{f}
 =  2\pi \int\limits_{0}^{\infty} \frac{\left[ \frac{J_{1}(\pi f)}{(\pi f)} -
   \frac{J_{1}(3 \pi f)}{(3 \pi f)}\right]^{2}}{f} df \nonumber\\
& = & 2\pi \left(\ln(3)-\frac{4}{9}\right) \approx 4.11 \nonumber
\end{eqnarray}    
\end{appendix}


\begin{thebibliography}{50}
\bibitem[Antoine \& Murenzi, 1996]{anto} Antoine J-P., Murenzi R., 1996,
Signal Processing 52, 259-281
\bibitem[Bensch et al.,{\em in. prep.},]{bensch}  Bensch F., Stutzki J., Ossenkopf
V.: Characterization of molecular cloud structure using the
$\Delta$-variance,
{\em in prep.}
\bibitem[Holschneider, 1995]{holschneider} Holschneider M., Wavelets - An Analysis Tool, Clarendon Press, Oxford 1995
\bibitem[Gill\- \& Hen\-riksen, \-1990]{gill} Gill A. G., Henriksen R. N.,
  1990, ApJ 365, L27-L30
\bibitem[Kramer et al., 1998]{kramer} Kramer C., Stutzki J., R\"ohrig
  R., Corneliussen U., 1998, A\&A, 329, 249
\bibitem[Langer et al., 1993]{langer} Langer W. D., Wilson R. W.,
Anderson C. H., 1993, ApJ 408, L45-L48 
\bibitem[Ossenkopf et al., {\em in prep.}]{ossk} Ossenkopf V., Stutzki
J., Bensch F.: Molecular cloud structure analysis and radiative transfer, {\em in prep.}
\bibitem[Pando \& Fang, 1998]{pando} Pando J., Fang L-Z., 1998,
  Phys. Rev. E 57, 3593-3601 
\bibitem[Stutzki et al, 1998]{stutzki} Stutzki J.,  Bensch F.,
  Heithausen H., Ossenkopf V., Zielinsky  M., 1998,  A\&A 336, 697-720 
\end{thebibliography}
\end{document}